\documentstyle[12pt]{article}
\def\centeron#1#2{{\setbox0=\hbox{#1}\setbox1=\hbox{#2}\ifdim
\wd1>\wd0\kern.5\wd1\kern-.5\wd0\fi
\copy0\kern-.5\wd0\kern-.5\wd1\copy1\ifdim\wd0>\wd1
\kern.5\wd0\kern-.5\wd1\fi}}
\def \ltap{\centeron{\raise.35ex\hbox{$<$}}{\lower.65ex\hbox{$\sim$}}}
\def \gtap{\centeron{\raise.35ex\hbox{$>$}}{\lower.65ex\hbox{$\sim$}}}
\def \gsim{\mathrel{\gtap}}
\def \lsim{\mathrel{\ltap}}
\begin{document}
\begin{titlepage}

\begin{flushright}
   hep-ph/9412381 \\
   MSUHEP--41222 \\
   May 1, 1995 \\
\end{flushright}

\begin{center}
  {\large \bf  Two Gluon Exchange Model Predictions \\
                for Double Pomeron Jet Production  }
  \vskip 0.30in
  {\bf Jon Pumplin }
\vskip 0.1in
Physics and Astronomy Department \\
Michigan State University\\
East Lansing MI 48824, U.S.A.

\end{center}

\vskip 0.5in

\begin{abstract}
I extend the two gluon exchange picture of elastic scattering,
known as the {\it Low-Nussinov} or {\it subtractive quark} model,
to predict cross sections for double pomeron exchange processes.
In particular, I calculate $p \bar{p} \to p \bar{p} q \bar{q}$
where the $q \bar{q}$ partons will appear as jets separated from
the final $p$ and $\bar{p}$ by large gaps in rapidity.
The predicted cross section is large enough that this process
should be observable at the Fermilab Tevatron and at the Large
Hadron Collider.  It can be distinguished from the background
of ordinary jet production by an absence of particles produced
in the gap regions.
\end{abstract}

\end{titlepage}

\section {Introduction}
\label{sec:intro}
The exchange of a gluon is the simplest interaction between
two hadrons in QCD.  It corresponds, via $s$-channel unitarity,
to an elastic amplitude dominated by exchange of two gluons
in a state with vacuum quantum numbers --- in particular, a color
singlet.  This provides a simple model, known as the
Low-Nussinov model \cite{lownuss,subtractive}, for the pomeron
that governs diffractive scattering at high energy.  The energy
dependence of the model ($s^1$ in the amplitude) is close to the
observed behavior ($\sim s^{1.08}$\cite{s108,zuoz}), so the
picture is qualitatively reasonable.

In this paper, we extend the Low-Nussinov model to predict cross
sections for double pomeron exchange (DPE)
processes \cite{dpe,dpeexpt,dpeheavyq,dpehiggs}, which are
characterized by two large rapidity gaps \cite{gaps,mult}.
It has been suggested \cite{dpe} that these processes will be
observable at the Fermilab Tevatron, and it is important to try
to predict their cross sections.  DPE will also be an interesting
subject for study at the Large Hadron Collider
($\sqrt{s} \cong 14 \, {\rm TeV}\,$).

We will focus on $q \bar q$ jet production in $p \bar p$ scattering,
where the final state contains only the two jets and a $p$
and $\bar p$ that carry $\gsim \! 95 \%$ of their original momenta.
The final $p$ and $\bar p$ have transverse momenta
$\lsim \! 1 \, {\rm GeV}$, putting them too close to the beam
directions (pseudo-rapidity $|\eta| > 7$) to be seen with present
detectors.  The absence of particles produced outside the two jets
(in Lego variables $\eta$ and $\phi$), except in the region between
them because of soft QCD radiation, contrasts strikingly to ordinary
events --- especially those with a hard scattering --- so the
signature of DPE-produced jets will be unmistakable.  Meanwhile, the
hard scattering amplitude is under control in perturbative QCD, so
no new parameters are added to the Low-Nussinov picture to make the
prediction.

To calculate two gluon exchange, we need a model for the internal
color structure of the hadrons to which the two gluon system
couples.  In this paper, the scattering hadrons are taken to be
$q \bar q$ bound states of effective ``quarks'' that have spin zero
and couple to the hadrons by a point coupling.  One might of course
prefer $qqq$ for the wave function of a baryon, and additional
$q \bar q$
pairs and gluons are certainly present in wave functions for small
momentum transfer.  Our simple model may nevertheless be adequate,
since only the distribution of color in the hadron is significant
for the calculation, and that can be modelled correctly by adjusting
parameters to fit elastic scattering.  Indeed, only the distribution
of color as a function of impact parameter really matters since the
spin $1$ gluon ``sees'' quarks equally, regardless of their
longitudinal momentum.  This justifies the simplicity of using
spin $0$ quarks.  We will also try an exponentially falling model
for the wave function, which is more realistic.  The model
dependence will be assessed by comparing results obtained using the
two different wave function types, with a range of choices for their
parameters.

Higher-order effects such as interaction between the exchanged
gluons must be important at some level, and is evidenced by the
deviation from constant total cross sections as a function of
energy.  Interactions could
even build a rather conventional Regge trajectory for the pomeron,
with physical glueball states on it at positive $t$, as Landshoff
has emphasized recently \cite{zuoz}.  More-than-two gluon exchange
contributions are also not negligible, as can be estimated by
eikonalizing the two gluon amplitude \cite{subtractive}.  (The
contribution beginning with four gluon exchange has sometimes also
been called double pomeron exchange \cite{mueller}.  It must not
be confused with the definition of DPE used here.)  We will neglect
these effects both for elastic scattering and jet
production.  Fitting the model parameters to elastic scattering
should reduce the consequences of this approximation.

In spite of its simplicity, it is worthwhile to see what this
model has to say about jet production in double pomeron exchange,
which has not until now been calculated --- even though more exotic
DPE processes of heavy quark \cite{dpeheavyq} and
Higgs \cite{milana} production have been.  The calculation will
be presented in considerable detail to make clear how it could
be applied to other double-diffractive processes.  The model could
also be applied to hard scattering with the exchange of just one
soft pomeron, i.e., hard scattering in diffractive dissociation.
Examples for study would be $\gamma \, p \to q \bar q p$, which can
be observed at HERA and has been calculated in somewhat different
models \cite{diehl,levin,brodsky}; and single-diffractive production
of $W^\pm$ which has been looked for by CDF \cite{cdfW}.

An additional theoretical motivation for this work is that the
two gluon exchange picture provides an explicit model for a
``direct'', ``coherent'', or ``lossless''
contribution \cite{cfs,SoperBerera}, in which the full energy of
the pomeron is available for hard scattering.  From a theoretical
standpoint, such contributions are interesting because they violate
the QCD
factorization rules that have been established for inclusive
processes, and that are often assumed without proof for the
diffractive subset of final states \cite{IngelmanSchlein,pompdf}.
They appear as an effective ``super-hard'' term $\propto \delta(x-1)$
in the phenomenological parton distribution of the pomeron.
Suggestive experimental evidence for a coherent contribution has
been seen by UA8 \cite{UA8}.

\section {Elastic Scattering}
\label{sec:elastic}
The pomeron is believed to arise from diffractive
physics, i.e., to be an $s$-channel unitarity phenomenon.
We therefore want the imaginary part of the
two gluon exchange amplitude, which
can be calculated for elastic scattering
from the discontinuity illustrated in Fig.~1:
\begin{eqnarray}
{\cal M} &=& \, \frac{-8}{2 i(2 \pi)^4} \,
\int \frac{d^4 k}{(k^2 - m_g^{\, 2}) \,
[(k + p_1 - p_3)^2 - m_g^{\, 2}]} \nonumber \\
& &\times \;
D_{\mu \nu}(p_1, k \to p_3,  k + p_1 - p_3) \,
D_{\mu \nu}(p_2,-k \to p_4, -k + p_2 - p_4) \; .
\label{eq:calm}
\end{eqnarray}
Note that the imaginary part is conveniently found by
cutting the diagram into two pieces through the possible
physical intermediate states.  Alternatively, one could
calculate the amplitude as a Feynman diagram, in which
case there would be an additional diagram in which the
two gluons cross each other.  The real part would cancel
between these diagrams because the amplitude has even
signature and energy dependence $\propto s^1$.

Eq.~(\ref{eq:calm}) contains a factor $8$ from the sum over
gluon colors.  A finite mass $m_g$ is included in the gluon
propagators to suppress contributions from long distance,
as an approximation to color confinement.
(An alternative modification of the gluon propagator
is discussed in Ref. \cite{cudell}.)

Our model for the discontinuity of the gluon-hadron amplitude,
illustrated in Fig.~2, is
\begin{eqnarray}
& &D_{\mu \nu}(p_1,k \to p_3, k + p_1 - p_3) = \nonumber \\
\, \null \nonumber \\  
& &\frac{g^2 \, G^2 }{2} \int \frac{d^4 q_1}{(2 \pi)^4}
\; [2 \pi i \, \delta(q_1^{\, 2} - m^2)]
\; [2 \pi i \, \delta(q_2^{\, 2} - m^2)] \, \nonumber \\
& &\times
\left[\frac{(p_1 - q_1 + q_2)_\mu}{(p_1 - q_1)^2 - m^2} \,
 - \, \frac{(p_1 - q_2 + q_1)_\mu}{(p_1 - q_2)^2 - m^2}
\right] \, \nonumber \\
\, \null \nonumber \\  
& &\times
\left[\frac{(p_3 - q_1 + q_2)_\nu}{(p_3 - q_1)^2 - m^2} \,
 - \, \frac{(p_3 - q_2 + q_1)_\nu}{(p_3 - q_2)^2 - m^2} \,
\right]
\label{eq:Dmunu1}
\end{eqnarray}
where $q_2 = k + p_1 - p_3$, and $g$ and $G$ are couplings of the
scalar quark to a gluon and to the hadron.  An overall factor
$\frac{1}{2}$ from color is included, although it is not
actually significant because the coupling strength $g^2 \, G^2$
is taken as a free parameter of the model.

In view of the delta functions, Eq.~(\ref{eq:Dmunu1}) appears to
be a two-dimensional integral.  In the high energy limit,
however, one of the delta functions can be reserved to apply to
the $d^4 k$ integral in Eq.~(\ref{eq:calm}), leaving a
three-dimensional integral.
To see this, introduce the light-cone coordinates
$p_\pm = (p_0 \pm p_z)/\sqrt{2}$ and work in a Lorentz frame
such as the center of mass, where $p_{1+} \cong p_{3+}$ and
$p_{2-} \cong p_{4-}$ are large with
$s = (p_1 + p_2)^2 \cong 2 \, p_{1+} \, p_{2-} \,$.
Then $d^4 q = d^2 q_\perp \, dq_+ \, dq_- \, $.  Introduce
$x_1 = q_{1+}/p_{1+}$ and $x_2 = q_{2+}/p_{1+}$ and use
$\delta(q_1^{\, 2} - m^2) =
\delta(2 \, q_{1+} \, q_{1-} - q_{1\perp}^{\, 2} - m^2)$
to do the $q_{1-}$ integral.  The other delta function becomes
\begin{eqnarray}
\delta(q_2^{\, 2} - m^2) \cong
\frac{\delta(k_-)}{2 \, q_{2+}} \cong
\frac{\delta(p_1 \! \cdot \! k)}{x_2}
\end{eqnarray}
in the high energy limit and we obtain
\begin{eqnarray}
D_{\mu \nu}(p_1,k \to p_3, k + p_1 - p_3) =
(p_1)_\mu \; (p_1)_\nu \;
\delta(p_1 \! \cdot \! k) \;
T(\vec{k}_\perp, (\vec{p}_3 - \vec{p}_1)_\perp)
\label{eq:Dmunu2}
\end{eqnarray}
where
\begin{eqnarray}
T(\vec{k}_\perp, (\vec{p}_3 - \vec{p}_1)_\perp) &=&
 \frac{-g^2 \, G^2}{8 \, \pi^2} \,
\int_0^1 dx_1 \,
\int_0^1 dx_2 \;
x_1 \, x_2 \; \delta(1 - x_1 - x_2) \,
\int d^2 \vec{q}_{1 \, \perp} \, \nonumber \\
\, \null \nonumber \\  
& &\times \,
  \left[\frac{1}{(\vec{q}_{1\perp} - x_1 \, \vec{p}_{1\perp})^{2}
+ {\tilde m}^2}
\, - \, \frac{1}{(\vec{q}_{2\perp} - x_2 \, \vec{p}_{1\perp})^{2}
+ {\tilde m}^2} \right] \, \nonumber \\
\, \null \nonumber \\  
& &\times \,
\left[\frac{1}{(\vec{q}_{1\perp} - x_1 \, \vec{p}_{3\perp})^{2}
+ {\tilde m}^2}
\, -  \, \frac{1}{(\vec{q}_{2\perp} - x_2 \, \vec{p}_{3\perp})^{2}
+ {\tilde m}^2} \right] \; .
\label{eq:T1}
\end{eqnarray}
Here
$\vec{q}_{2 \, \perp} = (\vec{p}_1 + \vec{k} - \vec{q}_1)_\perp$
and ${\tilde m}^2 = m^2 - x_1 x_2 M^2$ with $M$ the mass of
the hadron and $m$ the mass of the ``quark''.  We have ignored
the differences between
$(p_1)_\mu \, (p_1)_\nu$,
$(p_1)_\mu \, (p_3)_\nu$ and
$(p_3)_\mu \, (p_3)_\nu$, which are non-leading in $s$.

The transverse momentum integrals can be carried out to obtain
\begin{eqnarray}
T(\vec{k}_\perp, \vec{\Delta}_\perp) =
F(\vec{k}_\perp, \vec{\Delta}_\perp) \, - \,
F(\vec{0}, \vec{\Delta}_\perp) \label{eq:T2}
\end{eqnarray}
where
\begin{eqnarray}
F(\vec{k}_\perp, \vec{\Delta}_\perp) &=&
\frac{g^2 \, G^2}{4 \, \pi^2} \,
\int_0^1 \frac{dx}{x \, (1-x)} \,
\int d^2\vec{q}_\perp \;
\left[\frac{\vec{q}_\perp^{\, 2} + m^2}{x \, (1-x)}
- M^2 \right]^{-1} \, \nonumber \\
& &\times \;
\left[\frac{(\vec{q}_\perp + \vec{k}_\perp
- x \vec{\Delta}_\perp)^{\, 2} + m^2}{x \, (1-x)}
- M^2 \right]^{-1}
\label{eq:F1a} \\
\, \null \nonumber \\  
&=&
\frac{g^2 \, G^2}{2 \, \pi} \,
\int_0^1 dx \, x \, (1-x) \;
\frac{1}{A B} \,
\log{\frac{B+1}{B-1}}
\label{eq:F1b}
\end{eqnarray}
with $A = (\vec{k}_\perp - x \vec{\Delta}_\perp)^{\, 2}$ and
$B = \sqrt{1 + 4 {\tilde m}^2/A \,}\, $.
The remaining integral over quark momentum fraction $x$ can be
done numerically by Gauss-Legendre integration.

A more realistic model can be made by replacing the two energy
denominator factors of the form $[X - M^2 ]^{-1}$ in
Eq.~(\ref{eq:F1a}) by exponentials $\propto e^{-\beta X}$.  This
leads to
\begin{eqnarray}
F(\vec{k}_\perp, \vec{\Delta}_\perp) =
N \int_0^1 dx \;
e^{-\beta \, [\,(\vec{k}_\perp - x \vec{\Delta}_\perp)^{\, 2}
\, + \, 4\, m^2\,] / 2 \, x \, (1-x)}
\label{eq:F3a}
\end{eqnarray}
where the three constants $\beta$, $m$, and the normalization
$N$ parametrize the wave function in place of the point-coupling
model parameters.  The exponential form mimics the fragility
of actual hadrons at low momentum transfer, which is displayed
by the approximately exponential fall-off of elastic amplitudes
at small $-t$.  The structure of Eq.~(\ref{eq:F3a}) can be seen
better by writing it as
\begin{eqnarray}
F(\vec{k}_\perp, \vec{\Delta}_\perp) =
N \int_0^1 dx \;
e^{-(\beta / 2) \, [\,
(k_1^{\, 2} + 4m^2) / x  \; + \;
(k_2^{\, 2} + 4m^2) / (1-x) \; - \;
\vec{\Delta}_\perp^{\, 2} \,]}
\label{eq:F3b}
\end{eqnarray}
where
$k_1^{\, 2} = \vec{k}_\perp^{\, 2}$ and
$k_2^{\, 2} = (\vec{\Delta}_\perp - \vec{k}_\perp)^{\, 2}$ are
the squared momentum
transfers carried by each gluon.  A saddle-point approximation
to Eq.~(\ref{eq:F3b}) is convenient for speeding up
numerical calculations:
\begin{eqnarray}
F(\vec{k}_\perp, \vec{\Delta}_\perp) \cong
N \, \sqrt{\frac{2 \pi a b}{\beta}} \;
(a+b)^{-2} \;
 e^{-(\beta/2) \, [\, (a+b)^2 - \vec{\Delta}_\perp^{\, 2} \, ]}
\label{eq:F4}
\end{eqnarray}
where
$a = \sqrt{k_1^{\, 2} + 4m^2}$ and
$b = \sqrt{k_2^{\, 2} + 4m^2} \,$.

Returning to Eqs.~(\ref{eq:calm})--(\ref{eq:Dmunu2}), we have
\begin{eqnarray}
{\cal M} = \, \frac{i \, s}{8 \, \pi^4} \,
\int  \,
\frac{d^2 \vec{k}_\perp \,
[T(\vec{k}_\perp, \vec{\Delta}_\perp) \,]^2}
{(\vec{k}_\perp^{\, 2} + m_g^{\, 2}) \,
[(\vec{k}_\perp - \vec{\Delta}_\perp)^{\, 2} + m_g^{\, 2}]}
\label{eq:calm2}
\end{eqnarray}
with $\vec{\Delta}_\perp^{\, 2} \cong -(p_1 - p_3)^2 = -t$
the momentum transfer.  The $\vec{k}_\perp$ integral can
be done numerically.

We have three models for the proton wave function, given by
Eqs.~(\ref{eq:F1b}), (\ref{eq:F3b}) and (\ref{eq:F4}), with the
latter two equivalent except in computational convenience.  On
physical grounds, we estimate $m = 0.5$ (or $0.3$) and
$m_g = 0.3$ (or $0.14$ or $1.0$) in ${\rm GeV} = 1$ units, with
the alternatives representing an estimate of the possible range.

The model amplitude is proportional to $s^1$ as expected for
spin 1 exchange.  It thus describes an energy-independent total
cross section and elastic $d\sigma / dt$.  To choose the
remaining two parameters in each model, we fit the elastic
and total cross sections to
$\sigma_{\rm tot} = 65 \, {\rm mb}$ and
$\sigma_{\rm el} / \sigma_{\rm tot} = 0.207$, which are
based on a fit \cite{BSW} to $p \bar p$ elastic scattering
data \cite{UA4} at $\sqrt{s} = 546 \, {\rm GeV}$.  The parameter
values are shown in Table~I.

The parameters of the point-coupling model (Eq.~(\ref{eq:F1b})) are
such that $2 m$ is very close to $M$.  This can be understood
using the uncertainty principle:  the large spatial extent of the
proton, which is responsible for the large elastic slope, is
reproduced in the model by the possibility of nearly on-shell
dissociation of the proton into its constituents.  A choice like
$m = M = m_g = 0.3$, suggested in preliminary work by
Berera \cite{berera} for example, would instead make
$\sigma_{\rm el} / \sigma_{\rm tot} = 0.87$, which is much too
large.  Equivalently, but independent of the normalization, it
would make the average elastic slope
$\sigma_{\rm tot}^2/16 \pi \sigma_{\rm el}$
equal to $3.8 \, {\rm GeV}^{-2}$ instead of
$16.0 \, {\rm GeV}^{-2}$.

The exponentially falling wave function models (Eqs.~(\ref{eq:F3b}),
(\ref{eq:F4})) reproduce the actual shape of $d\sigma / dt$ much
better than the point coupling model.  These two models are very
similar, since the second is just a saddle-point approximation to
the first, with parameters chosen to give the same
$\sigma_{\rm tot}$ and $\sigma_{\rm el}$. Our smallest choice
$m_g = 0.14$ begins to have a slope at $t=0$ that is too steep,
since the slope diverges in the limit $m_g  \to 0$.

The hadronic discontinuity modelled by Fig.~2 is actually a
function of 3 scalar variables, say
$\vec{k}_{1\perp}^{\, 2}$,
$\vec{k}_{2\perp}^{\, 2}$, and
$(\vec{k}_{1\perp} + \vec{k}_{2\perp})^2$.  Tuning the model
parameters to fit the $t$-dependence of elastic scattering may
therefore not be sufficient to determine it.  (The ambiguity
could be reduced by also fitting the electromagnetic
form factor \cite{subtractive}.)  However, we will find that
the predictions are not extremely sensitive to the form of the
model.

Some important aspects of Eq.~(\ref{eq:calm2}) are especially clear
in the forward direction, where it reduces to
\begin{eqnarray}
\sigma_{\rm tot} = \frac{1}{8 \, \pi^4} \,
\int
d^2 \vec{k}_\perp \left(\frac
{F(\vec{0}, \vec{0}) \, - \, F(\vec{k}_\perp, \vec{0})}
{\vec{k}_\perp^{\, 2} \, + \, m_g^{\, 2}} \right)^2
\label{eq:sigtot}
\end{eqnarray}
The factor $F(\vec{0}, \vec{0}) \, - \, F(\vec{k}_\perp, \vec{0})$
represents the response of the hadronic wave function to gluon
momentum transfers of $\vec{k}_\perp$ and $-\vec{k}_\perp \,$.
$F(\vec{0}, \vec{0})$ comes from the ``diagonal'' diagrams in
Fig.~2, where both gluons hit the same quark line so there is
no $\vec{k}_\perp$ dependence from the wave function.
$F(\vec{k}_\perp, \vec{0})$ comes from the ``off-diagonal''
diagrams in which the gluons hit different quarks.

Eq.~(\ref{eq:sigtot}) has two different momentum scales:  one
associated with the gluon propagator and one associated with the
hadronic wave function.  The overall dependence on $\vec{k}_\perp$
is set by the fit to elastic scattering, but the relative
contributions are not well determined.

First consider the parameters in Table~I for Eq.~(\ref{eq:F3b})
with the smallest assumed $m_g = 0.14 \cong m_\pi$, which
corresponds to relatively long-range color confinement.  We find
$F(\vec{0}, \vec{0}) \, - \, F(\vec{k}_\perp, \vec{0})
\approx F(\vec{0}, \vec{0}) \,
(1 - e^{-11 \vec{k}_\perp^{\, 2}})$.  With these parameters,
the cancellation between diagonal
and off-diagonal terms is extremely important:  omitting the
off-diagonal term would increase $\sigma_{\rm tot}$ by a
factor $> \! 5\,$.  Meanwhile, the gluon mass is rather
unimportant:  even setting it to zero would increase
$\sigma_{\rm tot}$ by only a factor of $1.6\,$.
To emphasize the importance of cancellation between
contributions in which the gluons couple to the same or to
different quarks, which follows from color-neutrality of
the complete hadron, this picture has been called the
{\it subtractive quark model} \cite{subtractive}, in contrast
to the pre-QCD additive quark point of view.

Now consider instead the parameters in Table~I for
Eq.~(\ref{eq:F3b}) with the largest assumed value
$m_g = 1.0 \, {\rm GeV}$, which corresponds to very short-range
confinement.  We find
$F(\vec{0}, \vec{0}) \, - \, F(\vec{k}_\perp, \vec{0})
\approx F(\vec{0}, \vec{0}) \,
(1 - e^{-15 \vec{k}_\perp^{\, 2}})$. With this choice of
parameters, the off-diagonal term is quite unimportant:  omitting
it would increase $\sigma_{\rm tot}$ by only $10 \, \%$.  This
point of view in which the diagonal terms are dominant corresponds
to the additive quark picture advocated by Donnachie and
Landshoff \cite{zuoz,landshoff}.

We will find that the DPE predictions are somewhat different
for the different choices of $m_g$, so DPE measurements might
eventually be used to decide the correct point of view.

\section{\protect DPE Production of $q \bar q$ Jets}
\label{sec:DPEqq}

Fig.~3 shows a natural extension of the Low-Nussinov model to
describe $q \bar q$ jet production in DPE.  These diagrams can
be expected to dominate all other contributions because the
hadronic discontinuities $D_{\mu \nu}$ contain phase-coherent
sums over physical states, like their counterparts in
elastic scattering.  As in the case of elastic scattering, a
diagrammatic calculation would include crossed graphs that
individually generate real parts which cancel in the sum.
The discontinuity method is simpler as well as being more
intuitively related to $s$-channel unitarity.

The absorptive part of the amplitude is
\goodbreak
\begin{eqnarray}
{\cal M}_{q \bar q} &=& \, \frac{i \, g^2}{2 (2 \pi)^4} \,
\int d^4 k \; \frac{1}
{(k^2 - m_g^{\, 2}) \,
 (k_1^{\, 2} - m_g^{\, 2}) \,
 (k_2^{\, 2} - m_g^{\, 2})}
\nonumber \\
& &\times \;
[
 D_{\beta \mu}(p_1, k \to p_3, k_1) \,
 D_{\beta \nu}(p_2,-k \to p_4, k_2) \, +
\nonumber \\
& & \quad \;
 D_{\mu \beta}(p_1,-k_1 \to p_3,-k) \,
 D_{\nu \beta}(p_2,-k_2 \to p_4, k)
]
\nonumber \\
& &\times \;
\bar{u}(p_5) \left[
\frac{\gamma_\mu \, \gamma \! \cdot \! (p_5-k_1) \, \gamma_\nu}
     {(p_5-k_1)^2} \, + \,
\frac{\gamma_\nu \, \gamma \! \cdot \! (p_5-k_2) \, \gamma_\mu}
     {(p_5-k_2)^2}
\right] v(p_6)
\label{eq:calmqq1}
\end{eqnarray}
where $k_1 = p_1 - p_3 + k$ and $k_2 = p_2 - p_4 - k$.  Note
that we use true spin 1/2 massless quarks here, in contrast
to the effective ``quarks'' used in the wave function model.
Eq.~(\ref{eq:Dmunu2}) reduces this to an integral over
transverse momentum, and the contributions from the two sets
of diagrams in Fig.~3 are equal in view of the symmetry
$T(\vec{k}_\perp, \vec{\Delta}_\perp) =
T(\vec{\Delta}_\perp - \vec{k}_\perp, \vec{\Delta}_\perp)$.
This leads to
\begin{eqnarray}
{\cal M}_{q \bar q} &=&
\frac{i \, g^2}{16 \, \pi^4} \,
\int d^2 \vec{k}_\perp \, f(\vec{k}_\perp) \, A_{q \bar q}
\label{eq:calmqq2} \\
\, \null \nonumber \\  
f(\vec{k}_\perp) &=&
\frac{
T( \vec{k}_\perp, \vec{p}_{3\perp}) \;
T(-\vec{k}_\perp, \vec{p}_{4\perp})}
{
(\vec{k}_{ \perp}^{\, 2} + m_g^{\, 2}) \,
(\vec{k}_{1\perp}^{\, 2} + m_g^{\, 2}) \,
(\vec{k}_{2\perp}^{\, 2} + m_g^{\, 2})
}
\label{eq:calmqqf} \\
\, \null \nonumber \\  
A_{q \bar q} &=&
\bar{u}(p_5) \left[
\frac{\gamma \! \cdot \! p_1 \,
      \gamma \! \cdot \! (p_5-k_1) \,
      \gamma \! \cdot \! p_2}
     {(p_5-k_1)^2}  \, + \,
\frac{\gamma \! \cdot \! p_2 \,
      \gamma \! \cdot \! (p_5-k_2) \,
      \gamma \! \cdot \! p_1}
     {(p_5-k_2)^2}
\right] v(p_6)
\label{eq:Aqq}
\end{eqnarray}
where we set $\vec{p}_{1\perp} = \vec{p}_{2\perp} = 0$ for
incoming particles in the $\pm \, \hat{z}$ direction.

To compute $|{\cal M}_{q \bar q}|^2$, first compute
$\sum A_{q \bar q}^*(k^\prime) \, A_{q \bar q}(k)$ where the
sum is over $q$ and $\bar q$ helicities and $k^\prime$ is an
integration variable independent from $k$.  Neglect
non-leading powers in $s$ by dropping
$p_{1-}$, $k_{1-}$, $p_{2+}$, $k_{2+} \, $.  Let
\begin{eqnarray}
\vec{Q}_\perp = (\vec{p}_{5\perp} - \vec{p}_{6\perp})/2 \; ,
\label{eq:Qperp}
\end{eqnarray}
where $\vert \vec{Q}_{\perp}\vert$ is essentially the
transverse momentum of each jet since
$\vec{p}_{5\perp} + \vec{p}_{6\perp} =
-\vec{p}_{3\perp} - \vec{p}_{4\perp}$, with
$\vert \vec{p}_{3\perp}\vert$ and
$\vert \vec{p}_{4\perp}\vert$ limited to $\lsim \! 1 \, {\rm GeV}$
by the proton wave function.  The transverse momentum in the
loop integration is also limited by the proton wave function,
so  $Q_\perp$ is large compared to all other transverse momenta.
Keeping only the leading power in $Q_\perp$ gives
\begin{eqnarray}
\sum A_{q \bar q}^*(k^\prime) \, A_{q \bar q}(k) =
\frac{40 \, s^2}{3 \, Q_\perp^{\, 4} \, \cosh^4 \delta}
\, [a(k^\prime) \, a(k) \, \cosh^2 \delta
 \, + \,
 b(k^\prime) \, b(k) \, \sinh^2 \delta \, ]
\label{eq:SumAA}
\end{eqnarray}
where
\begin{eqnarray}
\delta = (y_5 - y_6)/2
\label{eq:delta}
\end{eqnarray}
and
\begin{eqnarray}
a(k) &=& (k_1)_x \, (k_2)_y  + (k_1)_y \, (k_2)_x \label{eq:aq} \\
b(k) &=& (k_1)_x \, (k_2)_x  - (k_1)_y \, (k_2)_y \label{eq:bq}
\end{eqnarray}
with
$\vec{k}_{1\perp} =   \vec{k}_{\perp} - \vec{p}_{3\perp} $,
$\vec{k}_{2\perp} = - \vec{k}_{\perp} - \vec{p}_{4\perp} $,
and
$\vec{Q}_\perp$ taken to be in the $\hat{x}$ direction.
Eq.~(\ref{eq:SumAA}) includes a factor of $\frac{16}{3}$ from
color and a factor of $5$ to sum over the quark jet flavors
$d$, $u$, $s$, $c$, $b$.  (Double-diffractive {\it top}
production will be a welcome newcomer at LHC.)

The cross section is
\begin{eqnarray}
\frac{d \sigma}{
d^2 \vec{p}_{3\perp} \,
d^2 \vec{p}_{4\perp} \,
d^2 \vec{Q}_{\perp} \,
dy_5 \, dy_6} \, = \,
\frac{
\vert {\cal M}_{q \bar q} \vert^2
}{
2^{12} \, \pi^8 \, s^2
}
\label{eq:dsigqq1}
\end{eqnarray}
where $y_5$ and $y_6$ are the rapidities of the two jets.
Both terms in Eq.~(\ref{eq:SumAA}) contain a function of
$\vec{k}_\perp^{\, \prime}$ times a function of $\vec{k}_\perp$.
Their contributions to $\vert {\cal M}_{q \bar q} \vert^2$ can
therefore be computed as absolute squares of integrals over
$\vec{k}_\perp$:
\begin{eqnarray}
\frac{d \sigma}{
d^2 \vec{p}_{3\perp} \,
d^2 \vec{p}_{4\perp} \,
d^2 \vec{Q}_{\perp} \,
dy_5 \, dy_6} &=&
\frac{\alpha_s^{\, 2}}{Q_\perp^{\, 4} \, \cosh^4 \delta}
\, (c_a \, \cosh^2 \delta + c_b \, \sinh^2 \delta)
\label{eq:dsigqq2}
\end{eqnarray}
where
\begin{eqnarray}
c_a &=&
\frac{10 }
{3 \, (2 \pi)^{14}} \,
\left\vert \, \int d^2 \vec{k}_\perp \,
f(\vec{k}_\perp) \, a(\vec{k}_\perp) \, \right\vert^2
\label{eq:caint} \\
c_b &=&
\frac{10}
{3 \, (2 \pi)^{14}} \,
\left\vert \, \int d^2 \vec{k}_\perp \,
f(\vec{k}_\perp) \, b(\vec{k}_\perp) \, \right\vert^2
\; .
\label{eq:cbint}
\end{eqnarray}
These results can also be obtained by calculating the individual
$q \bar q$ helicity amplitudes.  The cross sections are equal for
helicities
$(+\!1/2 \; -\!1/2)$ and
$(-\!1/2 \; +\!1/2)$, and zero for
$(+\!1/2 \; +\!1/2)$ and
$(-\!1/2 \; -\!1/2)$.

The cross section is independent of overall energy $s\,$.
The dependence on jet transverse momentum is the usual dimensional
$Q_\perp^{-4}$.  The dependence on
$\delta = (y_5 - y_6)/2$ is such that the two jets are usually
separated by $\lsim \! 2$ units of rapidity.  For large $\delta$,
the cross section falls as $e^{-|y_5 - y_6|}$ which is dictated
by Regge arguments for spin 1/2 exchange.  Similarly, there is no
dependence on the average rapidity $y_{\rm ave} = (y_5 + y_6)/2$
of the jet pair because the gluons have spin $1$.

In the special case where both leading particles have zero
transverse momentum $\vec{p}_{3\perp} = \vec{p}_{4\perp} = 0$,
the cross section is found to be zero as a result of the
azimuthal angle integrations in
Eqs.~(\ref{eq:caint})--(\ref{eq:cbint}).  This implies strong
correlations between the transverse momenta of the leading
particles.  A ``Regge factorization'' assumption, whereby the
cross section is a product of factors for emission of pomerons
by the fast particles times a cross section for two-pomeron
scattering, would be incorrect.  It also implies that it is
dangerous to estimate the DPE cross section on the basis of
the pure forward direction, as is done in somewhat different
models for heavy quark \cite{dpeheavyq} and Higgs \cite{milana}
production.

Integrating over the transverse momenta of both
quasi-elastically scattered $p$ and $\bar p$, since these
cannot be observed in current experiments, gives
\begin{eqnarray}
\frac{d \sigma}{
d^2 \vec{Q}_{\perp} \,
dy_5 \, dy_6} \, = \,
\frac{\alpha_s^{\, 2}}{Q_\perp^{\, 4} \, \cosh^4 \delta} \,
(C_a \, \cosh^2 \delta + C_b \, \sinh^2 \delta) \; .
\label{eq:dsigqq4}
\end{eqnarray}
When $c_a$ and $c_b$ are integrated over the azimuthal angles
of $\vec{p}_{3\perp}$ and $\vec{p}_{4\perp}$, they are found
to become equal.
{\it Hence $C_a = C_b$ in Eq.~(\ref{eq:dsigqq4}).}  Results are
shown in Table~I for our various choices of proton wave function
and gluon mass.
We find $C_a = C_b \approx 1.2 \times 10^{-3}$ with an uncertainty
up or down of a factor of $2.5 \,$.  Somewhat smaller results than
that are found for the rather extreme choice $m_g = 1.0 \,$.

The large $Q_\perp$ limit in Eqs.~(\ref{eq:SumAA})--(\ref{eq:bq})
was computed with the help of Mathematica \cite{mathematica}.  It
is interesting to compare it with production of {\it spin zero}
quarks, which is simple enough to work out by hand as follows.  The
spin $1/2$ factors in Eq.~(\ref{eq:calmqq1}) are replaced by the
likewise gauge-invariant form
\begin{eqnarray}
\frac
{(k_1 - 2p_5)_\mu \, (k_2 - 2p_6)_\nu }
{(k_1-p_5)^2} \, + \,
\frac
{(k_1 - 2p_6)_\mu \, (k_2 - 2p_5)_\nu}
{(k_1-p_6)^2} \, + \,
2 \delta_{\mu,\nu} \; .
\label{eq:spin0a}
\end{eqnarray}
Eq.~(\ref{eq:Aqq}) is replaced by
\begin{eqnarray}
A_{{\rm Spin }\, 0} &=& \frac
{s \, (1 - \alpha \beta)}
{(1 + \alpha) \, (1 + \beta)}
\label{eq:spin0b}
\end{eqnarray}
where
\begin{eqnarray}
\alpha &=&
(\vec{k}_1 - \vec{p}_5)_\perp^2 / 2 \, p_{5 -} \, p_{6 +} \nonumber \\
\beta &=&
(\vec{k}_1 - \vec{p}_6)_\perp^2 / 2 \, p_{5 +} \, p_{6 -}
\label{eq:spin0c}
\end{eqnarray}
It suffices to approximate
$\alpha \cong e^{2 \delta}$ and
$\beta \cong e^{-2 \delta}$
$\Rightarrow (1 + \alpha) \, (1 + \beta) \cong 4 \cosh^2 \delta$
in the denominator.  Terms of order $1/Q_\perp^2$ must be kept in
the numerator because $\alpha \beta$ is close to $1$, leading to
\begin{eqnarray}
A_{{\rm Spin }\, 0}
&=& \frac{s}{2 \, Q_\perp^2 \, \cosh^2 \delta} \,
\left(
2 \,
\hat Q_\perp \! \cdot \! \vec{k}_{1\perp} \;
\hat Q_\perp \! \cdot \! \vec{k}_{2\perp} \; - \;
\vec{k}_{1\perp} \! \cdot \! \vec{k}_{2\perp}
\right) \\
&=&
\frac{s}{2 \, Q_\perp^2 \, \cosh^2 \delta} \,
(k_{1x} \, k_{2x} \, - \, k_{1y} \, k_{2y})
\label{eq:spin0d}
\end{eqnarray}
where the final form  is for $\vec{Q}_\perp$ in the $\hat x$
direction.  The dependence on
$\vec{k}_{1\perp}$ and
$\vec{k}_{2\perp}$ is the same as in Eq.~(\ref{eq:bq}), so the
cross section again goes to zero in the double forward limit
$\vec{p}_{3\perp} = \vec{p}_{4\perp} = 0$.  The dependence of
the cross section on rapidity is $(\cosh \delta)^{-4}$, which
falls as $e^{-2 | y_5 - y_6|}$ for large separation as required
by Regge theory for spin $0$ exchange.

\section{Experimental Considerations}
\label{sec:experiment}
A measurement that could be made with the CDF or {D\O} detectors
at Fermilab (${\bar p}p$ at $\sqrt{s} = 1800 \, {\rm GeV}$) would
require two jets in the central region of pseudo-rapidity,
say $|\eta_5|$, $|\eta_6| < 1.5 \,$.  Defining the jets using a
cone radius of $0.7$ would leave regions of at least
$2.2 < |\eta| < 4.2$ in the two ``end-cap'' parts of the detector,
to observe the {\it absence} of produced hadrons that distinguishes
DPE from ordinary hard scattering.

The fraction of longitudinal momentum retained by the forward
incident proton is
$X_1 \cong
p_{3+} / p_{1+} \cong
1 - (p_{5+} + p_{6+})/p_{1+} \cong
1 - (e^{\eta_5} + e^{\eta_6}) Q_\perp / \sqrt{s} \,$.  Similarly,
$X_2 \cong
1 - (e^{-\eta_5} + e^{-\eta_6}) Q_\perp / \sqrt{s} \,$ is the
fraction of momentum retained by the backward anti-proton.
Requiring $X_1$, $X_2 > 0.95$ defines the DPE region as
\begin{eqnarray}
|\eta_5| &<& 1.5 \nonumber \\
|\eta_6| &<& 1.5 \nonumber \\
e^{ \eta_5} + e^{ \eta_6} &<&
0.05 \, \sqrt{s}/Q_\perp \nonumber \\
e^{-\eta_5} + e^{-\eta_6} &<&
0.05 \, \sqrt{s}/Q_\perp \; .
\label{eq:kine}
\end{eqnarray}

The predicted DPE cross section for $q \bar q$ jets is calculated by
integrating Eq.~(\ref{eq:dsigqq4}) over the region defined
by Eq.~(\ref{eq:kine}).  In doing this, I take
$\alpha_s(Q^2) = 12 \pi / (23 \, \ln Q^2/\Lambda^2)$
with $\Lambda = 0.2 \, {\rm GeV}$ and
$Q^2 = Q_\perp^{\, 2} / 4$.  The result is
\begin{eqnarray}
\sigma = \left\{
\begin{array}{l}
(3.516 \, C_a + 0.632 \, C_b) \, \mu b \quad
\mbox{for jets with } Q_\perp > 10 \, {\rm GeV/c} \\
(0.246 \, C_a + 0.040 \, C_b) \, \mu b \quad
\mbox{for jets with } Q_\perp > 20 \, {\rm GeV/c} \\
(0.022 \, C_a + 0.002 \, C_b) \, \mu b \quad
\mbox{for jets with } Q_\perp > 30 \, {\rm GeV/c} \; .
\end{array}
\right.
\label{eq:results1}
\end{eqnarray}
Using $C_a = C_b$ and taking $C_a = 1.2 \times 10^{-3}$ as a typical
estimate from Table~I gives
\begin{eqnarray}
\sigma = \left\{
\begin{array}{l}
4.98 \, {\rm nb} \quad
\mbox{for jets with } Q_\perp > 10 \, {\rm GeV/c} \\
0.34 \, {\rm nb} \quad
\mbox{for jets with } Q_\perp > 20 \, {\rm GeV/c} \\
0.03 \, {\rm nb} \quad
\mbox{for jets with } Q_\perp > 30 \, {\rm GeV/c} \; .
\end{array}
\right.
\label{eq:results2}
\end{eqnarray}
These predictions are
uncertain by a factor of 2--3 due to the model dependence
indicated by the spread of values for $C_a$.  The final predicted
cross sections will be larger because the contribution from gluon
jet production is yet to be calculated.

One might expect the cross section to be reduced by the following
``$t_{\rm min}$'' effect.  The four-momentum transfer to the
leading proton is
\begin{eqnarray}
t_1 = (p_1 - p_3)^2 = -[\, \vec{p}_{3\perp}^{\, 2}
\, + \, (1 - X_1)^2 \, m_p^2 \,]/X_1
\label{eq:t}
\end{eqnarray}
which becomes $-\vec{p}_{3\perp}^{\, 2}$ in the $X_1 \to 1$ limit
that is assumed in our calculation.  To correct for this
approximation, the predicted cross section should
be reduced by a factor
$\approx \! e^{B \, t_{\rm min}}$ where
$t_{\rm min} = t_1 + \vec{p}_{3\perp}^{\, 2} < 0$ and
$B \sim 16 \, {\rm GeV}^{-2}$ based on elastic scattering.
A similar factor would be expected for the anti-proton.
However, this effect is found to be small enough to neglect in
the region $X_1$, $X_2 > 0.95\,$.

Ordinary hard scattering generates a background to DPE
that I estimate using a {\footnotesize HERWIG} QCD Monte
Carlo simulation \cite{herwig} in the manner described in
Ref. \cite{mult}.  The predicted cross section
for 2 jets, each with $Q_\perp > 10 \, {\rm GeV}$, in the DPE
region defined by Eq.~(\ref{eq:kine}) is
$30 \, {\rm \mu b}$.  This cross section is nearly
4 orders of magnitude larger than the signal.  It is also
nearly $10^{-3}$ of the entire minimum
bias cross section, making it much too large to permit
experiments to trigger on every such event.

Imposing a rapidity gap condition on {\it one side},
by requiring zero particles of transverse momentum
$> 0.2 \, {\rm GeV}$ in the range
${\rm Max}(\eta_5, \eta_6) + 0.7 < \eta < 4.2$, reduces the
background by a factor $1/700$.  This makes it small enough
to permit triggering on all such ``single-gap'' jet events.

Imposing a rapidity gap condition on {\it both sides} by
requiring the regions
$-4.2 < \eta < {\rm Min}(\eta_5, \eta_6) - 0.7$ and
${\rm Max}(\eta_5, \eta_6) + 0.7 < \eta < 4.2$
empty of particles with $p_\perp > 0.2 \, {\rm GeV}$ leads to
a {\footnotesize HERWIG}-predicted background cross section of
$1.0 \, {\rm nb}$.

Our predicted cross section for quark jets alone is
a factor of 5 larger
than this background, so the DPE signal should show up
clearly as an ``extra'' contribution at zero
multiplicity in the particle multiplicity distribution
for the gap regions of two-jet events.  As a further
test of the model, the DPE region
could be tightened to $|\eta_5|$, $|\eta_6| < 1.0$ or
$< 0.5$, which would increase the minimum rapidity gaps
from $2.0$ to $2.5$ or $3.0 \,$.  This would very strongly
decrease the background from zero-multiplicity fluctuations
of ordinary jet production.  Of course, it would be
better to extend the observed gap regions to larger
$|\eta|$, or still better to detect the leading
$p$ and $\bar p\,$; but
those options require additions to the detectors.

\section{Conclusion}
\label{sec:conclusion}
We have combined the two gluon exchange model of the
pomeron with leading-order perturbative QCD for hard
scattering to predict cross sections for
$p \bar p \to p \bar p q \bar q$.
The process shown in Fig.~3 gives the dominant contribution
due to phase-coherence of the sums over intermediate states
represented by the hadronic discontinuities in Fig.~2.

A similar calculation of $p \bar p \to p \bar p gg$ is in
progress.  It is somewhat more complicated because many more
diagrams make up the appropriate gauge-invariant set.  The
only anticipated difference from the $q \bar q$ result is
that large rapidity separations will be possible between the
jets, as a result of having spin $1$ exchange in place
of spin $1/2$ between the jets.  This could be observable at
the LHC, but only
by means of detectors with a wider coverage in pseudorapidity
than those proposed so far.  It cannot be observed at the
Tevatron energy because jets with a large rapidity separation
would have too large an invariant mass to be produced in the
DPE region.

Our calculation resembles other QCD predictions, in that it
contains a long distance scale non-perturbative part (the
hadronic discontinuity which is related to a wave function),
and a short distance scale (high-$Q^2$) part that is calculated
perturbatively.  It differs from other predictions, however,
in that the non-perturbative part has been obtained by fitting
to low-$Q^2$ elastic scattering rather than
to a different high-$Q^2$ process; and in that there is no
factorization theorem to guarantee success.

A special feature of this exclusive DPE process is that, unlike
rapidity gaps created by other color singlet exchanges, there
is presumably no additional ``survival probability'' factor needed
to account for gaps that are filled in by incidental exchanges of
color, e.g., due to additional soft gluons exchanged between the
incident beam particles.  This is like elastic scattering itself.
Of course, there will be some suppression due to the fact that
soft particles from the jets can spread widely from the nominal
jet axes.  The effect of such particles can be reduced somewhat
by defining gaps as an absence of particles above some threshold
like $p_\perp > 0.2 \, {\rm GeV/c}$ \cite{mult}.

The experimental signature of our process is two hadronic jets
separated by $\lsim \! 2$ units of rapidity, and back-to-back in
azimuthal angle.  The final $p$ and $\bar p$ are at such small
angles with respect to the beam directions as to be undetectable
in present experiments.  Installing ``Roman Pot'' detectors to
cover the very small angle region would be valuable because it
would easily eliminate all backgrounds to DPE, and because there
are interesting correlations predicted between the transverse
momenta of the two leading particles relative to each other and
relative to the plane of the jets.  In particular, the predicted
cross section vanishes when both leading particles are at zero
transverse momentum.  This strongly contradicts a naive
assumption of Regge factorization.

The predicted cross section for the $q \bar q$ process
alone is $\sim 5 \, {\rm nb}$.  This is large enough to be
studied easily at the Tevatron and at the eventual Large Hadron
Collider at CERN.
To make the study, it will be necessary to have an
experimental trigger for the rapidity gap signature on at
least one side of the detector.  It will also be
necessary to use sufficiently low luminosity running
that the rapidity gaps are not filled in by particles
from additional $p \bar p$ collisions that occur during the
same beam crossing.

The production of jets discussed here is only one of many
possible DPE processes, since $gg \to q \bar q$ could
be replaced by any other hard scattering with a two gluon
initial state.  Some suggestions are given in
Ref. \cite{dpe}.  A further dramatic possibility would be
DPE production of a Higgs boson \cite{dpehiggs,milana}.

Some preliminary work on the subject of jet production
in DPE was presented at the
Fermilab Small-x Workshop \cite{berera,smallxjp}.

\section*{Acknowledgments}
I wish to thank my colleagues in the CTEQ collaboration,
especially J. Collins, H. Weerts, and S. Kuhlmann, for
discussions.  I also thank A. Berera, I. Balitsky, and
V. Braun for useful discussions.
I thank D. Zeppenfeld for pointing out the need to include
the partially-disconnected discontinuities in Fig. 3, which
were omitted in an earlier version of this paper.

\newpage
\begin{table}
\begin{center}Table I \\
Parameters of the model and predicted $C_a$
\end{center}
\begin{center}
\begin{tabular}{||l|c|c|c|c||c||}
\hline
\multicolumn{1}{||c|}{Model} &
\multicolumn{1}{|c|}{$m$} &
\multicolumn{1}{c|}{$m_g$} &
\multicolumn{2}{c||}{Other Parameters} &
\multicolumn{1}{c||}{$C_a = C_b$} \\
\hline
\hline
Eq.~(\ref{eq:F1b}) & $0.5$ & $0.14$ &
$M = 0.9818$ & $g^2 G^2 = 2.67$   & $2.7 \times 10^{-3}$ \\

Eq.~(\ref{eq:F1b}) & $0.3$ & $0.14$ &
$M = 0.5638$ & $g^2 G^2 = 1.976$  & $2.2 \times 10^{-3}$ \\
\hline
Eq.~(\ref{eq:F1b}) & $0.5$ & $0.30$ &
$M = 0.9868$ & $g^2 G^2 = 2.795$  & $1.2 \times 10^{-3}$ \\

Eq.~(\ref{eq:F1b}) & $0.3$ & $0.30$ &
$M = 0.5741$ & $g^2 G^2 = 2.011$  & $9.4 \times 10^{-4}$ \\
\hline
Eq.~(\ref{eq:F1b}) & $0.5$ & $1.0$ &
$M = 0.9916$ & $g^2 G^2 = 4.671$  & $6.7 \times 10^{-5}$ \\

Eq.~(\ref{eq:F1b}) & $0.3$ & $1.0$ &
$M = 0.5840$ & $g^2 G^2 = 3.285$  & $7.0 \times 10^{-5}$ \\
\hline
\hline
Eq.~(\ref{eq:F3b}) & $0.5$ & $0.14$ &
$\beta = 5.869$ & $N = 3.24 \times 10^7$  & $2.8 \times 10^{-3}$ \\

Eq.~(\ref{eq:F3b}) & $0.3$ & $0.14$ &
$\beta = 5.301$ & $N = 7.50 \times 10^3$  & $2.3 \times 10^{-3}$ \\

Eq.~(\ref{eq:F3b})& $0.0$ & $0.14$ &
$\beta = 2.378$ & $N = 7.23 \times 10^1 $ & $9.2 \times 10^{-4}$ \\
\hline
Eq.~(\ref{eq:F3b}) & $0.5$ & $0.30$ &
$\beta = 6.715$ & $N = 2.51 \times 10^8$  & $1.4 \times 10^{-3}$ \\

Eq.~(\ref{eq:F3b}) & $0.3$ & $0.30$ &
$\beta  = 6.222$ & $N = 2.05 \times 10^4$  & $1.1 \times 10^{-3}$ \\

Eq.~(\ref{eq:F3b}) & $0.0$ & $0.30$ &
$\beta  = 3.487$ & $N = 8.91 \times 10^1$  & $8.4 \times 10^{-4}$ \\
\hline
Eq.~(\ref{eq:F3b}) & $0.5$ & $1.0$ &
$\beta = 7.682$ & $N = 4.65 \times 10^9$  & $5.1 \times 10^{-5}$ \\

Eq.~(\ref{eq:F3b}) & $0.3$ & $1.0$ &
$\beta = 7.358$ & $N = 1.24 \times 10^5$  & $4.4 \times 10^{-5}$ \\

Eq.~(\ref{eq:F3b}) & $0.0$ & $1.0$ &
$\beta = 5.261$ & $N = 2.14 \times 10^2$  & $2.8 \times 10^{-4}$ \\
\hline
\hline
Eq.~(\ref{eq:F4}) & $0.5$ & $0.14$ &
$\beta  = 5.797$ & $N = 2.65 \times 10^7$ & $2.8 \times 10^{-3}$ \\

Eq.~(\ref{eq:F4}) & $0.3$ & $0.14$ &
$\beta  = 4.944$ & $N = 4.89 \times 10^3$ & $2.1 \times 10^{-3}$ \\
\hline
Eq.~(\ref{eq:F4}) & $0.5$ & $0.30$ &
$\beta = 6.659$ & $N = 2.13 \times 10^8$ & $1.4 \times 10^{-3}$ \\

Eq.~(\ref{eq:F4}) & $0.3$ & $0.30$ &
$\beta  = 5.901$ & $N = 1.40 \times 10^4$ & $1.0 \times 10^{-3}$ \\
\hline
Eq.~(\ref{eq:F4}) & $0.5$ & $1.0$ &
$\beta = 7.638$ & $N = 4.06 \times 10^9$ & $4.9 \times 10^{-5}$ \\

Eq.~(\ref{eq:F4}) & $0.3$ & $1.0$ &
$\beta = 7.098$ & $N = 9.04 \times 10^4$ & $4.1 \times 10^{-5}$ \\
\hline
\hline
\end{tabular}
\label{table1}
\end{center}
\end{table}


\clearpage
\section*{Figure Captions}

\begin{enumerate}

\item
Two gluon exchange (``Low-Nussinov'') model for the elastic amplitude.
The dashed line denotes an $s$-channel discontinuity.

\item
Quark model for the hadronic discontinuity in Fig.~1.

\item
Two gluon exchange model for Double Pomeron Exchange production
of $q \bar q$ jets.

\end{enumerate}


\newpage
\begin{thebibliography}{99}
\bibitem{lownuss}
F. Low, Phys.\ Rev.\ {\bf D12}, 163 (1975);
S. Nussinov, Phys.\ Rev.\ Lett.\ {\bf 34}, 1286 (1975);
J. Gunion and D. Soper, Phys.\ Rev.\ {\bf D15}, 2617 (1977).

\bibitem{subtractive}
J. Pumplin and E. Lehman, Zeit.\ Phys.\ {\bf C9}, 25 (1981);
J. Pumplin, Phys.\ Rev.\ {\bf D28}, 2741 (1983).

\bibitem{s108}
A. Donnachie and P. V. Landshoff,
Phys.\ Lett.\ {\bf B296}, 227 (1992).

\bibitem{zuoz}
P.V. Landshoff, Talk given at Summer School on Hadronic Aspects
of Collider Physics, Zuoz, Switzerland, 23-31 Aug 1994.
e-Print hep-ph/9410250.

\bibitem{dpe}
J. Pumplin, Phys.\ Rev.\ {\bf D47}, 4820 (1993).

\bibitem{dpeexpt}
D. Joyce et al., Phys.\ Rev.\ {\bf D48}, 1943 (1993);
M.G. Albrow, {\it Double Pomeron Exchange from the ISR to the SSC},
Talk given at the Int. Conf. on Elastic and Diffractive Scattering,
Evanston, Ill., May 2-6, 1989.
Nucl.\ Phys.\ B, Proc. Suppl. 12, 291 (1990).

\bibitem{dpeheavyq}
A. Bia\l as and W. Szeremeta, Phys.\ Lett.\ {\bf B296}, 191 (1992);

W. Szeremeta, Acta Phys.\ Pol.\ {\bf B24}, 1159 (1993);

A. Bia\l as and R. Janik, Z.\ Phys.\ {\bf C62}, 487 (1994).

\bibitem{dpehiggs}
A. Schafer, O. Nachtmann and R. Schopf,
Phys.\ Lett.\ {\bf B249}, 331 (1990);

A. Bia\l as and P.V. Landshoff,
Phys.\ Lett.\ {\bf B256}, 540 (1991);

B. Muller and A. Schramm,
Nucl.\ Phys.\ {\bf A523}, 677 (1991);

R.S. Fletcher and T. Stelzer,
Phys.\ Rev.\ {\bf D48}, 5162 (1993);

D. Zeppenfeld, ``Rapidity Gap Processes at Future Hadron
Colliders'', Talk given at 23rd International Symposium on
Ultra-High Energy Multiparticle Phenomena, Aspen, CO,
12-17 Sep 1993.

\bibitem{gaps}
J.D. Bjorken,
Int.\ J.\ Mod.\ Phys.\ {\bf A7}, 4189 (1992);
Phys.\ Rev.\ {\bf D45}, 4077 (1992);
Phys.\ Rev.\ {\bf D47}, 101 (1992);

R.S. Fletcher, Phys.\ Lett.\ {\bf B320}, 373 (1994).

\bibitem{mult}
J. Pumplin, Phys.\ Rev.\ {\bf D50}, 6811 (1994).

\bibitem{mueller}
A.H. Mueller and B. Patel, Nucl.\ Phys.\ {\bf B425}, 471 (1994).

\bibitem{milana}
H.-J. Lu and J. Milana,
{\sl Exclusive Production of Higgs Bosons in Hadron Colliders}
(Maryland U. preprint DOE-ER-40762-041) June 1994.
e-Print hep-ph/9407206.

\bibitem{diehl}
M. Diehl, ``Diffractive Production of Dijets at HERA''
(Cambridge U. DAMTP-94-60) e-Print hep-ph/9407399.

\bibitem{levin}
E. Levin and M. Wusthoff, Phys.\ Rev.\ D50, 4306 (1994).

\bibitem{brodsky}
S. J. Brodsky, L. Frankfurt, J.F. Gunion, A.H. Mueller,
and M. Strikman, Phys.\ Rev.\ {\bf D50}, 3134 (1994).

\bibitem{cdfW}
K. Goulianos, Talk at the Workshop on Small-x and Diffractive
Physics at the Tevatron (Fermilab, Sept. 22-24, 1994).

\bibitem{cfs}
J.C. Collins, L. Frankfurt, and M. Strikman,
Phys.\ Lett.\ {\bf B307}, 161 (1993).

\bibitem{SoperBerera}
A. Berera and D.E. Soper, Phys.\ Rev.\ {\bf D50}, 4328 (1994).

\bibitem{IngelmanSchlein}
G. Ingelman and P. Schlein,
Phys.\ Lett.\ {\bf B152}, 256 (1985).

\bibitem{pompdf}
J. Collins, J. Huston, J. Pumplin, H. Weerts, and J. Whitmore,
Phys.\ Rev.\ {\bf D51}, 3182 (1995).

J. Bartels and G. Ingelman,
Phys.\ Lett.\ {\bf B235}, 175 (1990);

H.-J. Lu and J. Milana,
Phys.\ Lett.\ {\bf B313}, 234 (1993);

P. Bruni and G. Ingelman,
Phys.\ Lett.\ {\bf B311}, 317 (1993);

N. Nikolaev and B. G. Zakharov,
Z. Phys.\ {\bf C53}, 331 (1992);

A. Donnachie and P.V. Landshoff,
Phys.\ Lett.\ {\bf 191B}, 309 (1987),
erratum: ibid.\ {\bf 198B}, 590 (1987).

\bibitem{UA8}
UA8 Collaboration, (A. Brandt et al.),
Phys.\ Lett.\ {\bf B297}, 417 (1992).

\bibitem{cudell}
M.B. Gay Ducati, F. Halzen and A.A. Natale,
Phys.\ Rev.\ {\bf D48}, 2324 (1993);

F. Halzen, G.I. Krein, and A.A. Natale,
Phys.\ Rev.\ {\bf D47}, 295 (1992);

J.R. Cudell and B.U. Nguyen,
Nucl.\ Phys.\ {\bf B420}, 669 (1994);

J.R. Cudell and B. Margolis,
Phys.\ Lett.\ {\bf B297}, 398 (1992).

\bibitem{BSW}
C. Bourrely, J. Soffer, and T. T. Wu,
Zeit.\ Phys.\ {\bf C37}, 369 (1988);
Phys.\ Lett.\ {\bf B252}, 287 (1990).
For a discussion of this fit, see
J. Pumplin, Phys.\ Lett.\ {\bf B276}, 517 (1992).

\bibitem{UA4}
UA4 Collaboration (M. Bozzo {\it et al.}),
Phys.\ Lett.\ {\bf B147}, 385 (1984); {\bf B147}, 392 (1984).

\bibitem{berera}
A. Berera, Talk at the Workshop on Small-x and Diffractive Physics
at the Tevatron (Fermilab, Sept. 22-24, 1994).

\bibitem{landshoff}
A. Donnachie and P.V. Landshoff,
Nucl.\ Phys.\ {\bf B244}, 322 (1984);
Nucl.\ Phys.\ {\bf B267}, 690 (1986);
P.V. Landshoff and O. Nachtmann,
Z.\ Phys.\ {\bf C35}, 405 (1987).

\bibitem{mathematica}
S. Wolfram, {\it Mathematica}, Addison-Wesley 1988, 1991.

\bibitem{herwig}
G.\ Marchesini and B.R.\ Webber,
Nucl.\ Phys.\ {\bf B310}, 461 (1988);
G.\ Abbiendi, I.G.\ Knowles, G.\ Marchesini, B.R.\ Webber,
M.H.\ Seymour and L.\ Stanco,
Comp.\ Phys.\ Comm.\ {\bf 67}, 465 (1992).

\bibitem{smallxjp}
J. Pumplin, Talk at the Workshop on Small-x and Diffractive Physics
at the Tevatron (Fermilab, Sept. 22-24, 1994).

\end{thebibliography}
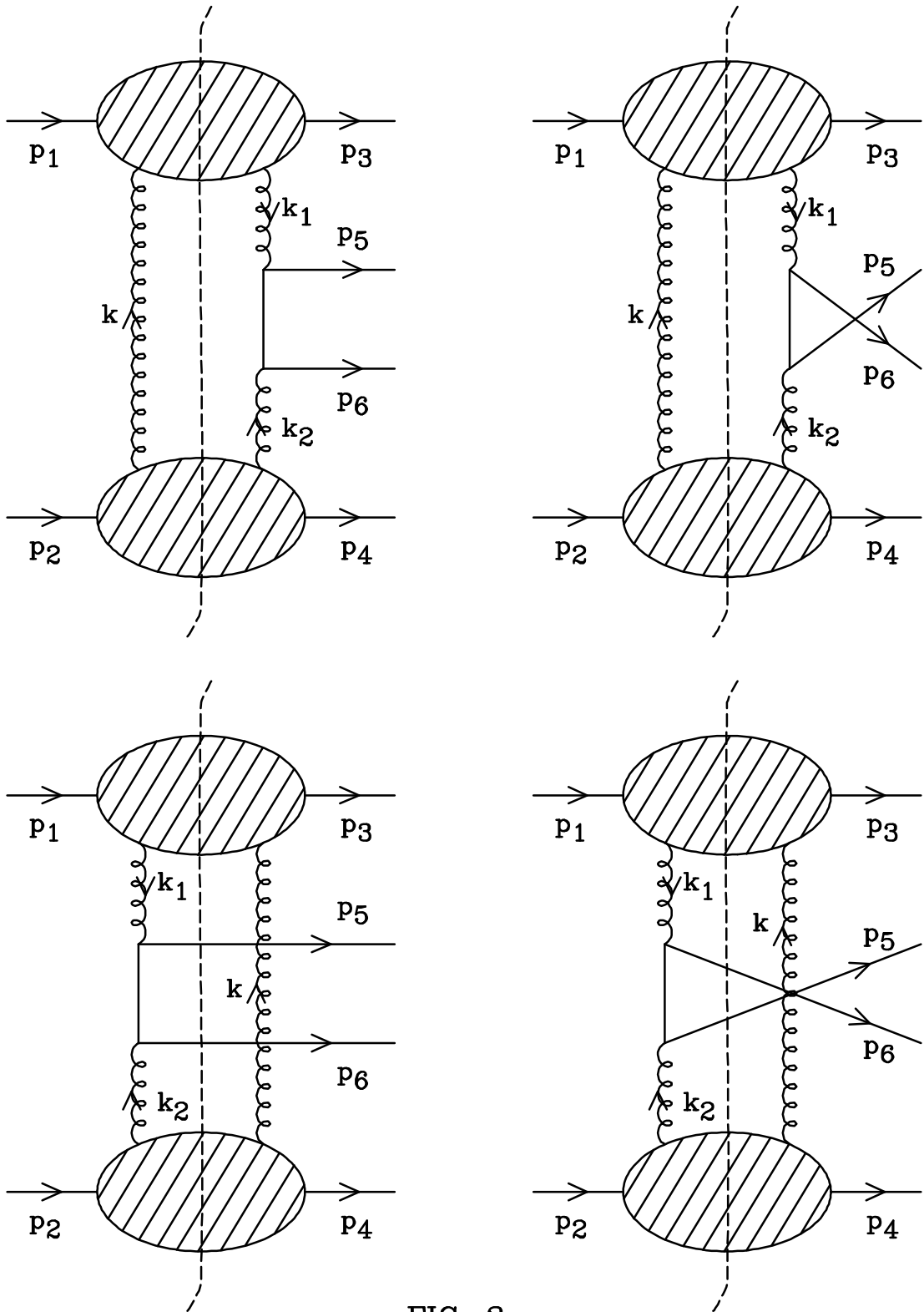
\end{document}